\documentclass[floatfix,twocolumn,superscriptaddress,showpacs,amsmath,amssymb,showpacs]{revtex4-2}

\usepackage{graphicx}
\usepackage{epstopdf}
\usepackage{epsfig}
\usepackage{epsf}
\usepackage{url}
\usepackage[USenglish]{babel}
\usepackage{hyperref}
\usepackage{verbatim}
\usepackage{xcolor}
\usepackage{natbib}
\usepackage{amsmath, nccmath}
\usepackage{dcolumn}
\usepackage{bm}
\usepackage{bbm}
\usepackage{lipsum}
\usepackage{makecell}

\begin{document}
\title{Strange diffusivity of incoherent metal in half-filled two-dimensional Hubbard model}
\author{Youngmin Eom}
\affiliation{Department of Physics and Chemistry, DGIST, Daegu 42988, Korea}
\author{Igor S. Tupitsyn}
\affiliation{Department of Physics, University of Massachusetts, Amherst, Massachusetts 01003, USA}
\author{Nikolay V. Prokof’ev}
\affiliation{Department of Physics, University of Massachusetts, Amherst, Massachusetts 01003, USA}
\author{Boris Svistunov}
\affiliation{Department of Physics, University of Massachusetts, Amherst, Massachusetts 01003, USA}
\author{Evgeny Kozik}
\affiliation{Department of Physics, Kings College London, Strand, London WC2R 2LS, United Kingdom}
\author{Aaram J. Kim}
\affiliation{Department of Physics and Chemistry, DGIST, Daegu 42988, Korea}

\begin{abstract}
We study charge transport across the metal-insulator crossover in the half-filled two-dimensional Hubbard model, with particular emphasis on precision control. The dynamic current-current correlation function is obtained directly in the thermodynamic limit, and the optical conductivity is extracted using numerical analytic continuation. To achieve this, we develop a multiscale approach: the non-perturbative low-frequency behavior is computed using the unbiased diagrammatic Monte Carlo technique, while the high-frequency physics is
captured via a self-consistent (semi-)analytic diagrammatic theory. We found that across a broad temperature range where the DC resistivity displays anomalous scaling, $\sim T^\alpha$ with $0<\alpha\lesssim 1$, the Nernst-Einstein relation implies the diffusion constant with the characteristic $\sim 1/\sqrt{T}$ ``strange metal" behavior. It was also revealed that the insulating regime is entered through a peculiar non-Fermi liquid state---which we call a \textit{Pseudogap Metal}---characterized by insulating charge compressibility coexisting with metallic transport. Diagrammatically, the high-temperature incoherent transport is captured by the dressed polarization bubble, whereas near the metal-insulator crossover, the effective interaction vertex between opposite-spin particles is responsible for transferring the Drude weight to a high-frequency continuum.

\end{abstract}

\maketitle


\textbf{Introduction}-
In incoherent metal, where the momentum relaxation rate is comparable to the temperature scale, the charge transport is dominated by collective diffusion. This state of matter cannot be described by the Fermi-liquid theory with well-defined quasi-particles;
it is phenomenologically known that the DC resistivity increases abnormally with temperature violating the $T^2$ scaling~\cite{Bruin2013}, and its magnitude can traverse the Mott-Ioffe-Regel (MIR) limit without hindrance~\cite{Ioffe1960,Mott1972,Gunnarsson2003,Hussey2004}.
Recently, several attempts have been made to unveil the governing principle of the incoherent metal from the diffusion viewpoint.
The universal bound on the charge diffusion constant $D$ is theoretically suggested by the hydrodynamic approach~\cite{Hartnoll2015}, while the ultracold atom experiments 
directly measure $D$ from the real-time charge relaxation process~\cite{Brown2019}.

The \textit{half-filled} $2d$ Hubbard model on the square lattice
\begin{equation}
	\hat{\mathcal{H}} = -t\sum^{}_{\langle ij\rangle\sigma}(\hat{c}^{\dagger}_{i\sigma}\hat{c}^{}_{j\sigma} + h.c.) + U\sum^{}_{i}\left(\hat{n}_{i\uparrow}-\frac{1}{2}\right)\left(\hat{n}_{i\downarrow} - \frac{1}{2}\right)~,
	\label{eqn:HM}
\end{equation}
with the hopping amplitude $t$ (set as the energy unit), Hubbard repulsion $U$, and the spin $\sigma$ fermionic annihilation (creation) operators $\hat{c}^{(\dagger)}_{i\sigma}$ on site $i$,
provides an ideal platform to investigate the charge diffusion process of the incoherent metal.
First, transport in this model is intrinsically anomalous due to the perfect nesting of the Fermi surface.
Enhanced charge susceptibility at the nesting momentum transfer $\mathbf{Q}=(\pi,\pi)$ results in a linear-in-$T$ scaling of the incoherent relaxation rate even in the weak-coupling regime~\cite{Lee1987,Virosztek1990,Kiely2021,Vucicevic2023}.
Second, particle-hole symmetry at half-filling completely decouples charge and heat carriers resulting in vanishing thermoelectric response~\cite{Beni1975}.
Consequently, one can access the charge diffusion constant
using charge compressibility $\kappa$ and DC conductivity $\sigma_{\text{DC}}$ via simple Nernst-Einstein relation 
\begin{equation}
	D = \sigma_{\text{DC}}/\kappa~.
	\label{eqn:NernstEinstein}
\end{equation}

\begin{figure}[b]
	\centering
	\includegraphics[width=0.95\columnwidth]{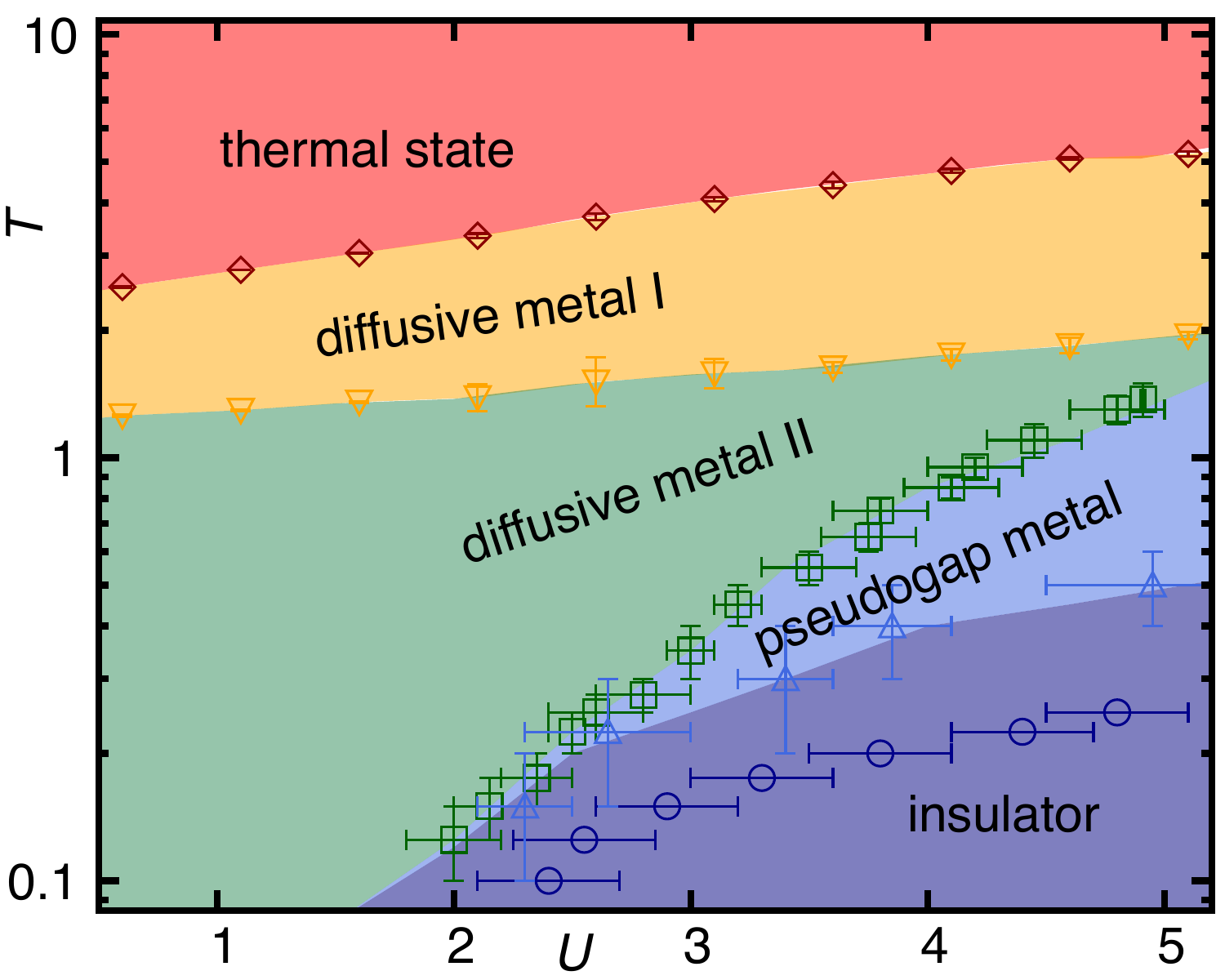}
	\caption{
		Various temperature regimes are classified by the DC resistivity, the compressibility, and the diffusion constant.
		The boundaries between the thermal state, Diffusive Metal I, Diffusive Metal II, and Pseudogap Metal (diamond, inverted triangle, and square symbols) are determined by the compressibility and diffusion constant, while the metal-to-insulator crossover line (blue triangles) marks the temperature of the $\rho_{\text{DC}}$ minimum for a given $U$.
		The spin-crossover temperature is indicated by (navy) circles~\cite{Kim2020}.
	}
	\label{fig:TUdiagram}
\end{figure}

\begin{figure*}[t]
	\centering
	\includegraphics[width=1.0\textwidth]{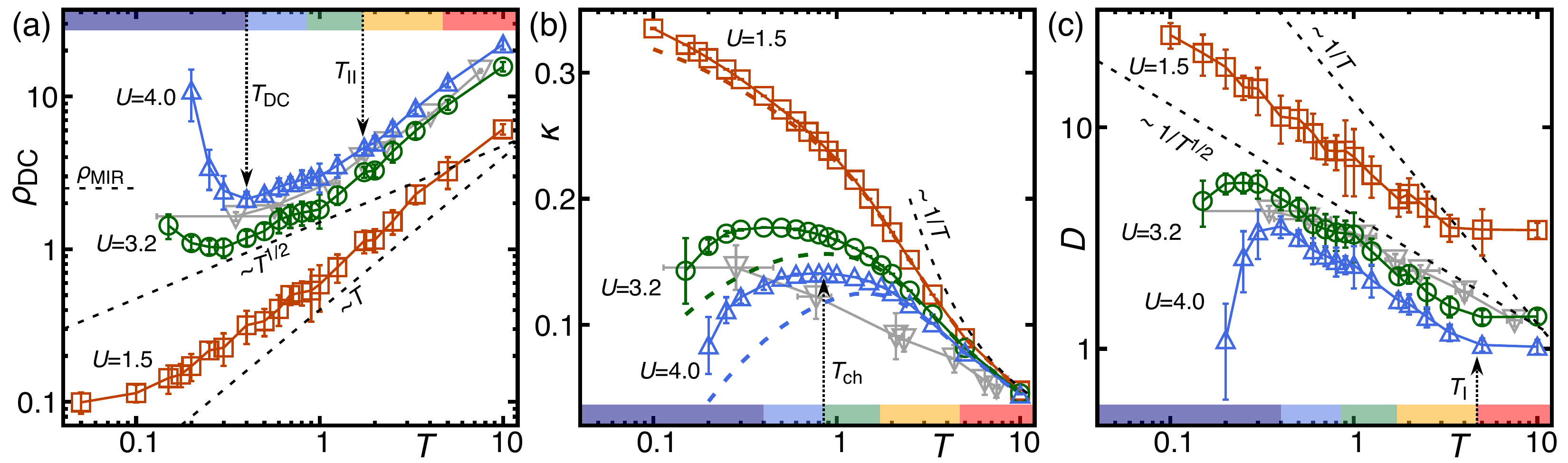}
	\caption{
		Nernst-Einstein decomposition of (a) the DC resistivity $\rho_{\text{DC}}$ into (b) the compressibility $\kappa$ and (c) the diffusion coefficient $D$ as functions of temperature for three different $U$ values: 1.5, 3.2, and 4.
		The inverted triangles (grey) for all panels present experimentally measured (or determined) value for the hole-doped system~\cite{Brown2019}.
		The MIR limit in panel (a), $\rho_{\text{MIR}}=\sqrt{2\pi}$~\cite{Gunnarsson2003,Kokalj2017}.
		In the panel (b), the first-order compressibility results
		are shown as the dashed lines, with colors matched to the corresponding full high-order results.
		The horizontal color bars exhibit different temperature regimes at $U=4$, whose crossover scales -- $T_{\mathrm{I}}, T_{\mathrm{II}}, T_{\mathrm{ch}}$, and $T_{\mathrm{DC}}$ -- are marked by dotted arrows. 
		For the definitions, see the main text.
	}
	\label{fig:EinsteinDecomposition}
\end{figure*}

Finally, the finite-temperature behavior of the half-filled $2d$ Hubbard model is relatively well understood. Not only have its single-particle~\cite{Simkovic2020} and two-particle (spin and charge) properties~\cite{Kim2020} been systematically studied, but its thermodynamic behavior~\cite{Connor2021} has also been investigated using numerically exact methods and cross-benchmarked by various state-of-the-art approaches at weaker coupling~\cite{Schaefer2021}. Despite the simple antiferromagnetic ground state arising from perfect Fermi surface nesting, previous studies revealed rich behavior at finite temperature. In this regime, where the true long-range order is suppressed because of the Mermin-Wagner theorem, thermal, spin, and charge fluctuations become strongly intertwined, giving rise to a transient non-Fermi liquid state.

\setlength{\tabcolsep}{1.0em} 
{\renewcommand{\arraystretch}{1.4}
\begin{table*}
	\centering
	\begin{tabular}{ccccc}
		\hline\hline
		Insulator & Pseudogap Metal (PGM) & Diffusive Metal II (DM2) & Diffusive Metal I (DM1) & Thermal State (TS)\\
		\hline
		$\frac{d\rho_{\text{DC}}}{dT}<0$ & $\rho_{\text{DC}}\sim T^{\alpha}$ & $\rho_{\text{DC}}\sim T^{\alpha}$ & $\rho_{\text{DC}}\sim T$  & $\rho_{\text{DC}}\sim T$\\
		$\frac{d\kappa}{dT}>0$ & $\frac{d\kappa}{dT}>0$ & $\frac{d\kappa}{dT}<0$ & $\kappa\sim 1/T^{(1/2+\delta)}$ & $\kappa\sim 1/T$ \\
		$\frac{dD}{dT}>0$ & $D\sim 1/\sqrt{T}$  & $D\sim 1/\sqrt{T}$ & $D\sim 1/T^{(1/2-\delta)}$ & $D\sim const.$\\
		\hline\hline
	\end{tabular}
	\caption{
		Summary of temperature dependence of $\rho_{\text{DC}}$, $\kappa$, and $D$~. The resistivity exponent of the Pseudogap Metal (Diffusive Metal II) has range of $0\lesssim\alpha\lesssim 0.5$  ($0.5\lesssim\alpha\lesssim 1$)~.
	}
	\label{tab:summary}
\end{table*}
}

In this paper, we analyze charge transport and corresponding diffusion process of the model (\ref{eqn:HM}) using the diagrammatic Monte Carlo method (DiagMC)~\cite{Prokofev1998,Kozik2010,VanHoucke2010}
tuned for stochastically sampling diagrammatic series of the charge and current correlation functions.
DiagMC evaluates the diagrammatic expansion to high order numerically exactly, enabling the evaluation of observables directly in the thermodynamic limit with controlled error bars.
We use the connected determinant Monte Carlo (CDet) algorithm~\cite{Rossi2017} to access high orders (up to order 10).
Across a wide range of temperatures featuring an anomalous scaling of the DC resistivity, the diffusion constant obtained via Eq.~(\ref{eqn:NernstEinstein}) displays very robust $\sim 1/\sqrt{T}$ scaling, which resembles the recent results of cold atom experiments at non-zero doping~\cite{Brown2019}.
At low temperature, we observe an intriguing state referred to here as a Pseudogap Metal with the mixed character of metallic transport and insulating
charge response.
Diagrammatically, this non-Fermi liquid state is characterized by considerable vertex corrections. The vertex component that couples currents of opposite spins is responsible for transferring the low-frequency Drude weight to the high-frequency continuum.

To obtain the optical conductivity $\sigma(\omega)$, the current autocorrelation function (See the supplementary materials (SM) Sec. I~\cite{supp}.)
\begin{align}
	\Lambda(i\omega_n) &= - 2t^2\int_{0}^{\beta}d\tau~e^{i\omega_n\tau}\sum_{i,\sigma\sigma'} \left\langle \mathcal{T}_{\tau} \hat{c}^{\dagger}_{i\sigma}(\tau)\hat{c}^{}_{i+\hat{x},\sigma}(\tau)\right.\nonumber\\
    &\times\left.\left(\hat{c}^{\dagger}_{0\sigma'}(0)\hat{c}^{}_{0+\hat{x},\sigma'}(0) - \hat{c}^{\dagger}_{0\sigma'}(0)\hat{c}^{}_{0-\hat{x},\sigma'}(0)\right)\right\rangle ~,
	\label{eqn:LambdaTwoTerms}
\end{align}
for the bosonic Matsubara frequency $\omega_n = 2n\pi/\beta$ with the inverse temperature $\beta$, needs to be continued numerically to the real frequency axis via the spectral representation, 
\begin{equation}
	\Lambda(i\omega_n)= \int_{-\infty}^{\infty}d\omega~K(i\omega_n,\omega)\sigma(\omega),
	\label{eqn:spectralRep}
\end{equation}
with the kernel $K(i\omega_n,\omega) = \omega^2/\pi(\omega_n^2+\omega^2)$.
Numerical analytic continuation (NAC) falls into the mathematically ill-defined class of Fredholm equations of the first kind, which generally leads to artifacts caused by the saw-tooth instability
even for accurate $\Lambda(i\omega_n)$ data due to the small 
eigenvalues of the kernel $K$. We employ the unbiased stochastic optimization with consistent constraints method (SOCC)~\cite{Mishchenko2000,Mishchenko2012,SOCC} and improve the accuracy by a multiscale treatment of $\Lambda(i\omega)$: The higher-frequency behavior, where the physics becomes more perturbative, is computed without statistical noise using a diagrammatic theory with self-consistent renormalization in the fermionic and three bosonic channels---the so-called Bold4 method introduced in Ref.~\cite{Simkovic2017}.

Although there has been significant recent progress on this notorious NAC problem via stochastic methods~\cite{Beach2004,Sandvik1998,Mishchenko2000,Fuchs2010,Mishchenko2012,SOCC}, sparse sampling~\cite{yoshimi2019,motoyama2022}, machine learning~\cite{Yoon2018,Fournier2020}, and pole estimations based on the Herglotz-Nevanlinna functions~\cite{Fei2021a,Fei2021b,Huang2023b,Zhang2024a,Zhang2024b} beside the classical approaches such as Pad\'e approximation~\cite{Baker1996,Schoett2016} and maximum entropy method~\cite{Bryan1990,Jarrell1996}, for smooth spectra $\sigma(\omega)$ that satisfy Eq.~(\ref{eqn:spectralRep}) within the error bars, all methods are equally legitimate; see, e.g., the discussion in Ref.~\cite{SOCC}.
The key challenge is not finding a solution (or hundreds of them!) but systematic errors on its characteristic features, such as locations and widths of maxima, minima, and $\sigma(0)$---the inferred DC conductivity. 
We define this error using a ``stretch test"~\cite{SOCC}, which quantifies how much the value can be shifted from ``as-smooth-as-possible" solution before $\sigma(\omega)$ develops the saw-tooth instability. We confirm that this estimate is consistent with the error bars produced by the SOCC method.
(For details on diagrammatic series convergence and the NAC protocol, see Sec. II and III in SM~\cite{supp}.)

\textbf{Summary of finite-temperature states}- Properties of the DC resistivity, $\rho_{\text{DC}}=1/\sigma(0)$, compressibility $\kappa$---directly computed by DiagMC from the charge correlation function~\cite{Kim2020}---and the diffusion constant $D$, which is inferred through the relation (\ref{eqn:NernstEinstein}),
partition the $T$-$U$ parameter plane into five different regimes shown in Fig.~\ref{fig:TUdiagram} (from high to low temperature): Thermal State (TS), Diffusive Metal I (DM1) and II (DM2), Pseudogap Metal (PGM), and Insulator. These distinct states are defined by a particular behavior of $\rho_{\text{DC}}$, $\kappa$, and $D$, summarized in Table~\ref{tab:summary}, and are delimited by the temperature crossover scales $T_{\mathrm{I}}$, $T_{\mathrm{II}}$, $T_{\mathrm{ch}}$, and $T_{\mathrm{DC}}$, as defined below and illustrated in Fig. \ref{fig:EinsteinDecomposition}.

Figure~\ref{fig:EinsteinDecomposition}(a) shows the temperature dependence of $\rho_{\text{DC}}$ for three representative $U$ values: 1.5, 3.2 and 4.
For all three $U$ values, the $T$-linear behavior at the ultrahigh temperatures (TS) persists down to the DM1, where the system is no longer described by the atomic limit. 
Further cooling reveals the underlying effects of the van Hove singularity in the compressibility ($T_{\textrm{II}}$) and drives the system into an even more anomalous transport regime---DM2 and later PGM---where the resistivity scales as $\sim T^\alpha$ with $0<\alpha< 1$~.
At the lower temperature bound of DM2, the compressibility marks the maximum and from that point on becomes insulator-like, entering the PGM state.
Finally, the metal-to-insulator crossover (MIC) intervenes, where $d\rho_{\mathrm{DC}}/dT$ changes sign. 

\textbf{$T$-linear resistivity extending from Thermal State to Diffusive Metal I}- 
Although the $T$-linear scaling of $\rho_{\text{DC}}$ appears in both TS and DM1, 
the temperature dependence of the compressibility and the diffusion constant differs qualitatively between these two regimes.
In TS, where the temperature is the largest energy scale in the system, the $T$-linear behavior of resistivity arises primarily from the corresponding change in the charge compressibility, while the diffusivity is almost constant.
These trends are well consistent with previous results from high-temperature expansion~\cite{Perepelitsky2016}. 
In Fig.~\ref{fig:EinsteinDecomposition}(b,c) for $T\gtrsim 5$, one can find the $1/T$ behavior in $\kappa$ with the almost saturated $D$.
The boundary of the TS ($T_\text{I}$), defined by the onset of varying diffusivity
\footnote{Technically, $T_\text{I}$ can also be defined by the condition $\partial^2 \kappa/\partial\log T^2=0$, in agreement with the definition in the main text within the error bars~\cite{supp}.},
increases with $U$ as shown in Fig.~\ref{fig:TUdiagram}~.

Below $T_\text{I}$, 
the diffusion constant grows as a power law upon cooling, while the compressibility increases more slowly than at ultrahigh temperatures.
The simultaneous increase in $D$ and the slowdown of the increase in $\kappa$ create a nontrivial balance that retains the $T$-linear resistivity away from ultrahigh temperatures.

\textbf{$1/\sqrt{T}$ diffusivity in a wide range of temperatures from Diffusive Metal I to Pseudogap Metal}- From DM1, the increase of the diffusion constant with cooling clearly follows $\sim 1/\sqrt{T}$ scaling, which hasn't been reported at the half-filling.
It is worth to emphasize that the recent cold-atom experiment~\cite{Brown2019} (and the finite-size calculation~\cite{Huang2019}) for the doped $2d$ Hubbard model also shows a power-law scaling of $D$ with similar exponents, $\sim T^{-0.6}$ [grey symbols in Fig.~\ref{fig:EinsteinDecomposition}(c)] in spite of the substantial doping level ($\langle n\rangle=0.82(2)$) and different interaction strength ($U=7.4(8)$).
Regarding the clear difference in the compressibility between the half-filled (this work) and doped (cold-atom experiment) case, it is surprising that the diffusion constant of two systems behaves in such a similar way. It is tempting to conjecture $D\sim 1/\sqrt{T}$ as a universal feature of incoherent transport.

\begin{figure}[b]
	\centering
	\includegraphics[width=1.0\columnwidth]{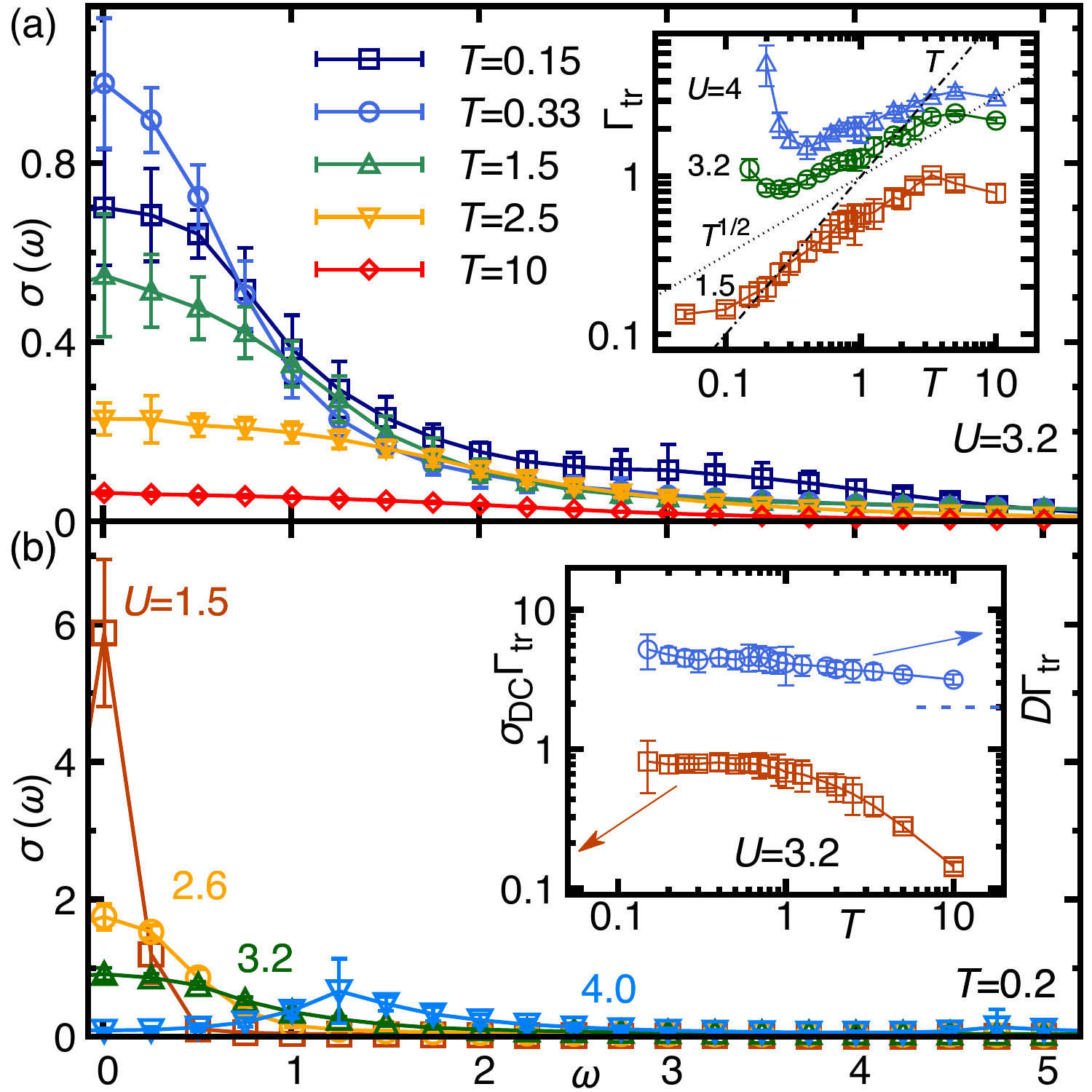}
	\caption{
		Optical conductivity $\sigma(\omega)$ for various (a) temperatures and (b) interaction strengths.
		The inset of panel (a) presents the momentum relaxation rate for $U=1.5, 3.2$ and $4$~.
		$A\Gamma_{\text{tr}}$ where $A=\sigma_{\text{DC}}$ or $D$, is shown in the inset of  panel (b), 
		and (blue) horizontal dashed line represents the limiting value at weak coupling and high temperature~\cite{Vucicevic2023}.
	}
	\label{fig:sigmaOmega}
\end{figure}

Interestingly, the $\sim 1/\sqrt{T}$ scaling of the diffusion constant persists all the way down to the MIC temperatures ($T_{\text{DC}}$) across a wide range of $U$ values. 
For example, at $U=4.0$, it spans the temperature window $0.4\lesssim T\lesssim 4.0$, covering DM1, DM2 and PGM in Fig.~\ref{fig:TUdiagram}, whereas the compressibility and the resistivity exhibit non-universal temperature dependencies.
The compressibility even changes the sign of its derivative $d\kappa/dT$ at $T_{\text{ch}}$ in this temperature window, becoming insulator-like and leading to the Pseudogap Metal
between $T_{\text{DC}}<T<T_{\text{ch}}$.

\textbf{Pseudogap metal}- The PGM region in Fig.~\ref{fig:TUdiagram} coincides with the regime where the one-particle self-energy was found to display a significant anisotropy~\cite{Simkovic2020}. In this regime, the spectral pseudogap develops first at the anti-nodal point in the Brillouin zone and proliferates toward the nodal point upon cooling, progressively suppressing the compressibility [Fig.~\ref{fig:EinsteinDecomposition}(b) below $T_{\mathrm{ch}}$].
Diagrammatically, such suppression can be captured by the strong scattering between particle-hole excitations near the anti-nodal point shown by the first-order charge vertex diagram.
The dashed lines in Fig.~\ref{fig:EinsteinDecomposition}(b) represent the first-order compressibility curves, which include the bubble and the first-order vertex diagrams.
The characteristic $\sim |\log T|$ increase in the particle-hole excitations (bubble diagram) near the anti-nodal point due to the van Hove singularity is compensated by the scattering process $\sim -U|\log T|^2$ (first-order diagram), e.g. around $T\sim 1.5$ for $U=4$.
At lower temperatures, this vertex correction eventually dominates, leading to the insulating charge response~\cite{supp}.
Compared to the first-order estimate, the full high-order crossover temperature $T_{\text{ch}}$ is quantitatively smaller, with $\sim 0.8$ for $U=4$ [Fig.~\ref{fig:EinsteinDecomposition}(b)].

Meanwhile, as signaled by Fig.~\ref{fig:sigmaOmega}(a) inset, the quasi-particle scattering rate near the remaining nodal points will decrease with cooling.
The reduced scattering rate 
can lead to an increase in the diffusion constant,
often estimated as $\sim \langle v^2\rangle\tau$~\cite{Kokalj2017}, where $\langle v^2\rangle$ is the mean-square quasi-particle velocity and $1/\tau$ is the scattering rate.  
Although the mean-square velocity is also suppressed due to the self-energy (we confirm that the quasi-particle weight estimated by $\kappa/\kappa_0$ monotonically decreases, not shown), the reduction in the scattering rate prevails and brings the overall increase in diffusivity. 
Since, around its maximum at $T_\text{ch}$, the compressibility varies weakly with temperature, the DC resistivity of the PGM, whose temperature dependence is mostly dominated by the diffusion constant, approximately scales $\sim \sqrt{T}$.

\textbf{Frequency dependent conductivity}- The full frequency dependence of the conductivity shown in Fig.~\ref{fig:sigmaOmega} provides further information on the finite energy excitations including the evolution of the central Drude peak.
Cooling from the ultrahigh temperature TS for $U=3.2$ and $T=10$ [red line in Fig.~\ref{fig:sigmaOmega}(a)], the height of the Drude peak is progressively enhanced while the momentum relaxation rate $\Gamma_{\text{tr}}$---estimated from the half width at half maximum (HWHM)---is nearly saturated.
As shown in the inset of Fig.~\ref{fig:sigmaOmega}(b), the saturated $\Gamma_{\text{tr}}$ is consistent with the high-temperature, weak-coupling limit where $D\Gamma_{\text{tr}}\rightarrow 2t^2$~\cite{Vucicevic2023}.

However, in the lower-temperature DM1 ($T=2.5$), $\Gamma_{\mathrm{tr}}$ begins to decrease while $\sigma_{\text{DC}}$ continues to increase with cooling.
The decrease in $\Gamma_{\text{tr}}$ turns out to be rather slow compared to $\sim 1/T$ increase in $\sigma_{\text{DC}}$, leading to overall increases of the Drude weight ($\propto \sigma_{\text{DC}}\Gamma_{\text{tr}}$) as shown in Fig.~\ref{fig:sigmaOmega}(b).
Especially, for $U=3.2$ and $4$, due to its slow decrease, the estimated $\Gamma_{\text{tr}}$ eventually becomes larger than the temperature scale signaling \textit{incoherent transport} in the DM1 and DM2
\footnote{Since no time scale is supposed to be larger than the temperature, in this case, the HWHM should be interpreted as a signal of the strong renormalization of other quantities.}.
Such trends in $\sigma_{\text{DC}}$ and $\Gamma_{\text{tr}}$ continue down to the DM2 ($T=1.5$) where the estimated Drude weight $\sigma_{\text{DC}}\Gamma_{\text{tr}}$ saturates at low temperatures ($T=0.33$).

As the system crosses over to the insulating regime with enhanced $\rho_{\text{DC}}$ ($T=0.15$), a noticeable high-energy continuum appears.
In this parameter regime, the strongly enhanced antiferromagnetic correlations were previously observed~\cite{Kim2020,Schaefer2021}.
As a function of $U$ for a low temperature $T=0.2$~[Fig.~\ref{fig:sigmaOmega}(b)], the central Drude peak monotonically decreases and eventually gives rise to the broad dip at $U=4$.

\begin{figure}[t]
	\centering
	\includegraphics[width=1.0\columnwidth]{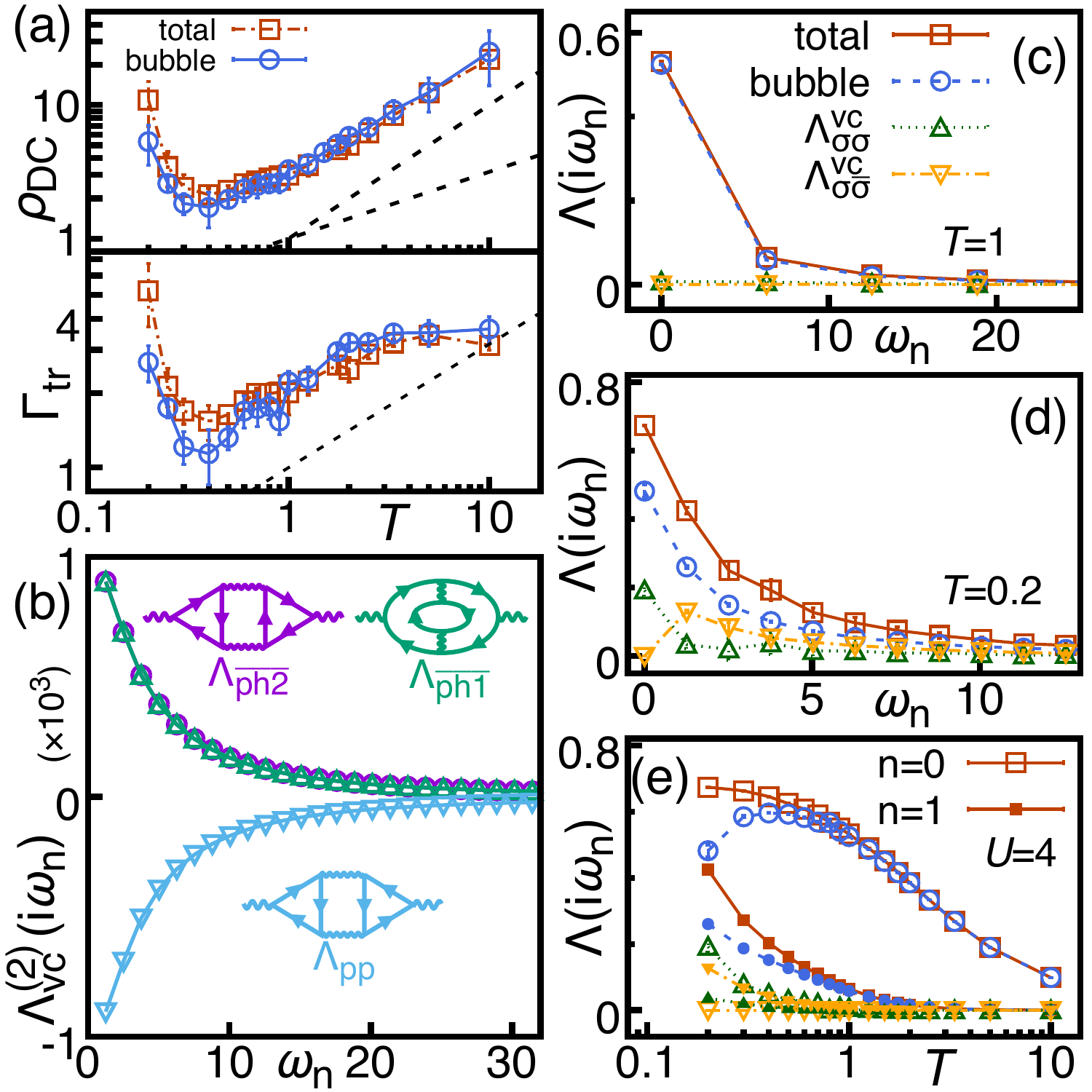}
	\caption{
		(a) The dressed bubble contribution to $\rho_{\text{DC}}$ and $\Gamma_{\text{tr}}$.
		(b) Exact second-order vertex diagrams: two opposite-spin vertices, $\Lambda_{\text{pp}}$ and $\Lambda_{\overline{\text{ph}2}}$ cancel perfectly, leaving only the same-spin vertex ($\Lambda_{\overline{\mathrm{ph1}}}$).
		The diagrammatic decomposition (bubble, same-spin vertex, and opposite-spin vertex) of the Matsubara current-current correlation function for (c) $T=1$ and (d) $0.2$ with the fixed $U=4$~.
		The panel (e) shows the temperature dependence of the current-current correlation function decomposed into the dressed bubble and the spin-dependent vertex corrections. 
		Open (solid) symbols denote the zeroth (first) Matsubara frequency results.
	}
	\label{fig:bubble}
\end{figure}
\textbf{Diagrammatic origin of the incoherent transport}- 
In order to track down the diagrammatic origin of the incoherent transport, we selectively extract the dressed bubble diagrams from the current-current correlation function, and further decompose the remaining vertex diagrams into spin-resolved components exploiting the perfect cancellation of certain vertex diagrams at the half filling (e.g. Fig.~\ref{fig:bubble}(b))~\cite{supp}.
Figure~\ref{fig:bubble}(a) shows the bubble contribution to the DC resistivity $\rho_{\text{DC}}^{\text{bubble}}$ and the corresponding momentum relaxation rate $\Gamma^{\text{bubble}}_{\text{tr}}$.
For the TS, DM1, and DM2, the bubble contributions of $\rho_{\text{DC}}^{\text{bubble}}$ and $\Gamma^{\text{bubble}}_{\text{tr}}$ are identical to the total values within the error bars, indicating negligible vertex corrections.
This small vertex contribution can also be confirmed at the level of $\Lambda(i\omega_n)$ on the Matsubara frequency in Fig.~\ref{fig:bubble}(c).

However, in the PGM, the total vertex contribution (sum of green and yellow lines in Fig.~\ref{fig:bubble}(e)) 
starts to increase, and eventually leads to the sizable difference between the bubble and total results in the insulating regime [Fig.~\ref{fig:bubble}(d)].
Without vertex contributions, $\rho_{\text{DC}}^{\text{bubble}}$ and $\Gamma^{\text{bubble}}_{\text{tr}}$ sizably underestimate the total values.
Such suppression and broadening of the Drude peak by the current vertex is qualitatively consistent with the results from the dynamical vertex approximation~\cite{Kauch2020} and the dual-$GW$ method~\cite{Dasari2025}.

In the PGM and Insulator regimes, where vertex contributions are significant, 
e.g. $U=4$ and $T=0.2$, the current-current correlation functions of different spin components play distinct roles.
Figure~\ref{fig:bubble}(d) shows the spin-resolved vertex corrections in the Matsubara current correlation functions: the same ($\Lambda^{vc}_{\sigma\sigma}$, green) and opposite ($\Lambda^{vc}_{\sigma\bar{\sigma}}$, yellow) spin components.
The same-spin component has a sizable value only for zero Matsubara frequency ($\omega_0$) whereas the opposite-spin one has a largest value at $\omega_1$ with vanishing $\omega_0$ component.
Based on the spectral representation [Eq.~(\ref{eqn:spectralRep})],
in addition to the bubble optical conductivity, the same-spin vertex adds sharp optical weight within $2\pi T$ window, while the opposite-spin counterpart \textit{redistributes} these weights, particularly in a way that suppresses the low-energy Drude peak.
In this way, the system can maintain an increasing total conductivity while depleting the central Drude peak as temperature decreases. 
If one considers only the bubble contributions, the total conductivity starts to decrease at a clearly higher temperature, e.g. around $T\sim 0.4$ for $U=4$~[open circles in Fig.~\ref{fig:bubble}(e)].

\textbf{Conclusion}-
We present a comprehensive picture of the charge transport in the prototypical $2d$ Hubbard model at the half-filling.
The numerically exact current autocorrelation function, analytically continued to real frequency, reveals the anomalous charge transport satisfying the defining properties of the strange metal, the $T$-linear DC resistivity and the violation of the MIR limit.
In particular, it is shown that such $T$-linearity is the product of a subtle balance between the compressibility and the diffusion constant.
As the resistivity exponent $\alpha$ continuously changes from one to smaller values with cooling, the underlying diffusion constant shows the robust $\sim 1/\sqrt{T}$ scaling down to the metal-to-insulator crossover, remarkably resembling to the one away from half-filling, directly measured in the ultracold atom experiment~\cite{Brown2019}.
This unexpected coincidence might imply universality of the scaling of the diffusion constant in the incoherent metallic transport,
and calls for further investigation to bridge these two separate parameter regimes.

As the temperature is lowered, even more anomalous transport 
appears in the Pseudogap Metal due to substantial vertex corrections.
Despite the strong scattering of particle–hole excitations near the anti-nodal point, which leads to the insulating compressibility, the reduced momentum relaxation rate near the nodal point ensures an overall metallic transport in the Pseudogap Metal.
The spin-dependent diagrammatic decomposition of the current-current correlator reveals a peculiar role of vertex corrections in the Pseudogap Metal and Insulator: the same-spin vertex contributes to the total spectral weight, whereas the opposite-spin vertex transfers the low-energy Drude weight to the high-energy continuum.

\textbf{Acknowledgements}- We thank A. Kauch, Y. W. Son, G.-Y. Guo, J. Tomczak, and G. Rohringer for helpful discussions.
A.J.K. and Y.M.E. acknowledge support from the DGIST Start-up Fund Program of the Ministry of Science and ICT (2024010026). 
E.K. was supported by EPSRC through Grant No. EP/X01245X/1.
I.T., B.S., and N.P.  acknowledge support from the National Science Foundation under Grant No. DMR-2335904.  
We also acknowledge the DGIST Supercomputing Big Data Center for the dedicated allocation of computing time.

\bibliography{ref_ref}

\pagebreak
\onecolumngrid
\begin{center}
\textbf{\large Supplemental Materials: Strange diffusivity of incoherent metal in half-filled two-dimensional Hubbard model}
\end{center}
\setcounter{equation}{0}
\setcounter{figure}{0}
\setcounter{table}{0}
\setcounter{page}{1}
\makeatletter
\renewcommand{\theequation}{S\arabic{equation}}
\renewcommand{\thefigure}{S\arabic{figure}}

\section{Determinant representation of (bubble) current-current correlation function}
The zero-momentum correlation function between the $x$-directional currents along the imaginary-time $\tau$ is defined as
\begin{align}
	\Lambda(\tau) &= \frac{1}{N_x}\langle \mathcal{T}_\tau\hat{J}^x(\tau)\hat{J}^x(0)\rangle~,\nonumber\\
	&= -\frac{t^2}{N_x}\sum^{}_{ij\sigma\sigma'}\left\langle \mathcal{T}_\tau\left(\hat{c}^{\dagger}_{i+\hat{x}\sigma}(\tau)\hat{c}^{}_{i\sigma}(\tau) - \hat{c}^{\dagger}_{i\sigma}(\tau)\hat{c}^{}_{i+\hat{x}\sigma}(\tau)\right)\left(\hat{c}^{\dagger}_{j+\hat{x}\sigma'}(0)\hat{c}^{}_{j\sigma'}(0) - \hat{c}^{\dagger}_{j\sigma'}(0)\hat{c}^{}_{j+\hat{x}\sigma'}(0)\right)\right\rangle~,
	\label{eqn:Lambda}
\end{align}
where $\hat{c}^{}_{i\sigma}$ ($\hat{c}^{\dagger}_{i\sigma}$) represents the annihilation (creation) operator of the spin-$\sigma$ fermion at site $i$~. 
Here, $i+\hat{x}$ denotes the neighboring site of $i$ shifted to the positive $x$ direction, and $N_x$ is the number of lattice sites.
Using translation and inversion symmetries, Eq.~(\ref{eqn:Lambda}) can be simplified as
\begin{equation}
	\Lambda(\tau) = -2t^2 \sum_{i} \sum_{\sigma\sigma'} \left[\left\langle \mathcal{T}_{\tau} \hat{c}^{\dagger}_{i\sigma}(\tau)\hat{c}^{}_{i+\hat{x},\sigma}(\tau)\left(\hat{c}^{\dagger}_{0\sigma'}(0)\hat{c}^{}_{0+\hat{x},\sigma'}(0) - \hat{c}^{\dagger}_{0\sigma'}(0)\hat{c}^{}_{0-\hat{x},\sigma'}(0)\right)\right\rangle \right]~.
	\label{eqn:LambdaTwoTerms}
\end{equation}
For a given site $i$ and spin species $\sigma$ and $\sigma'$, there are two terms in Eq.~(\ref{eqn:LambdaTwoTerms}), each of which corresponds to the sum of the connected diagrams. 
One can calculate those terms from a \textit{single} determinant (or a product of block-diagonal determinants for two different spins) of the noninteracting Green's function matrix.
Explicitly, in the diagrammatic expansion of Eq.~(\ref{eqn:LambdaTwoTerms}), the sum of all possible (both connected and disconnected) $k$-th order diagrams for a given set of internal space-time vertices $V = \{v_1, \cdots, v_k\}$ and external vertices $E = \{v_0, v\}$ can be written as below; for the same-spin current correlation ($\sigma'=\sigma$)
\begin{equation}
	\left|
	\begin{array}{ccc|c|c}
	    G_{\sigma}^0(v_1,v_1)-\frac{1}{2} & \cdots & G_{\sigma}^0(v_1,v_k) & G_{\sigma}^0(v_1,v_0)  & G_{\sigma}^0(v_1,v)\\
        \vdots & \ddots & \vdots & \vdots & \vdots \\
	G_{\sigma}^0(v_k,v_1) & \cdots & G_{\sigma}^0(v_k,v_k)-\frac{1}{2} & G_{\sigma}^0(v_k,v_0)  & G_{\sigma}^0(v_k,v)\\
	\hline
	\makecell{G_{\sigma}^0(v_0+\hat{x},v_1) \\- G_{\sigma}^0(v_0-\hat{x},v_1)} & \cdots & \makecell{G_{\sigma}^0(v_0+\hat{x},v_k) \\- G_{\sigma}^0(v_0-\hat{x},v_k)} & \makecell{G_{\sigma}^0(v_0+\hat{x},v_0) \\- G_{\sigma}^0(v_0-\hat{x},v_0)} & \makecell{G_{\sigma}^0(v_0+\hat{x},v) \\- G_{\sigma}^0(v_0-\hat{x},v)}\\
	\hline
	G_{\sigma}^0(v+\hat{x},v_1) & \cdots & G_{\sigma}^0(v+\hat{x},v_k) & G_{\sigma}^0(v+\hat{x},v_0)  & G_{\sigma}^0(v+\hat{x},v)
	\end{array}
	\right|
	\times
	\left|
	\begin{array}{ccc}
	    G_{\bar{\sigma}}^0(v_1,v_1)-\frac{1}{2} & \cdots & G_{\bar{\sigma}}^0(v_1,v_k)\\
        \vdots & \ddots & \vdots\\
	G_{\bar{\sigma}}^0(v_k,v_1) & \cdots & G_{\bar{\sigma}}^0(v_k,v_k)-\frac{1}{2}\\
	\end{array}
	\right|~,
	\label{eqn:sameSpin}
\end{equation}
while for the opposite spins ($\sigma'=\bar{\sigma}$)
\begin{equation}
	\left|
	\begin{array}{ccc|c}
	    G_{\sigma}^0(v_1,v_1)-\frac{1}{2} & \cdots & G_{\sigma}^0(v_1,v_k)  & G_{\sigma}^0(v_1,v)\\
        \vdots & \ddots & \vdots & \vdots \\
	G_{\sigma}^0(v_k,v_1) & \cdots & G_{\sigma}^0(v_k,v_k)-\frac{1}{2} & G_{\sigma}^0(v_k,v)\\
	\hline
	G_{\sigma}^0(v+\hat{x},v_1) & \cdots & G_{\sigma}^0(v+\hat{x},v_k) & G_{\sigma}^0(v+\hat{x},v)
	\end{array}
	\right|
	\times
	\left|
	\begin{array}{ccc|c}
	    G_{\bar{\sigma}}^0(v_1,v_1)-\frac{1}{2} & \cdots & G_{\bar{\sigma}}^0(v_1,v_k) & G_{\bar{\sigma}}^0(v_1,v_0)\\
        \vdots & \ddots & \vdots & \vdots\\
	G_{\bar{\sigma}}^0(v_k,v_1) & \cdots & G_{\bar{\sigma}}^0(v_k,v_k)-\frac{1}{2} & G_{\bar{\sigma}}^0(v_k,v_0)\\
	\hline
	\makecell{G_{\bar{\sigma}}^0(v_0+\hat{x},v_1) \\- G_{\bar{\sigma}}^0(v_0-\hat{x},v_1)} & \cdots & \makecell{G_{\bar{\sigma}}^0(v_0+\hat{x},v_k) \\- G_{\bar{\sigma}}^0(v_0-\hat{x},v_k)} & \makecell{G_{\bar{\sigma}}^0(v_0+\hat{x},v_0) \\ -G_{\bar{\sigma}}^0(v_0-\hat{x},v_0)}
	\end{array}
	\right|~.
	\label{eqn:oppoSpin}
\end{equation}
Here, $v_j+\hat{x}$ denotes the space-time coordinate $v_j$ whose spatial component is shifted to the positive $x$ direction.
After computing the determinants of all possible subset matrices of Eq.~(\ref{eqn:sameSpin}) and (\ref{eqn:oppoSpin}), the disconnected diagrams can be recursively subtracted in a standard protocol~\cite{Rossi2017,Simkovic2022}.

In order to access the \textit{bubble-only} contributions out of the full same-spin correlator [Eq.~(\ref{eqn:sameSpin})], we split the set of internal vertices $V=\{v_1,\cdots,v_k\}$ into two disjoint sets, $V_1=\{v_1,\cdots,v_{k_1}\}$ and $V_2=\{v_{k_1+1},\cdots,v_k\}$, and separately compute the Green's functions connecting $v_0$ and $v$ dressed by either $V_1$ or $V_2$.
Specifically, the Green's function $G_1$ from $v_0$ to $v+\hat{x}$ dressed by $V_1$ is the connected part of the determinant below
\begin{equation}
	\left|
	\begin{array}{ccc|c}
		G_{\sigma}^0(v_1,v_1)-\frac{1}{2} & \cdots & G_{\sigma}^0(v_1,v_{k_1}) & G_{\sigma}^0(v_1,v_0)\\
        \vdots & \ddots & \vdots & \vdots \\
	G_{\sigma}^0(v_{k_1},v_1) & \cdots & G_{\sigma}^0(v_{k_1},v_{k_1})-\frac{1}{2} & G_{\sigma}^0(v_{k_1},v_0)\\
	\hline
	G_{\sigma}^0(v+\hat{x},v_1) & \cdots & G_{\sigma}^0(v+\hat{x},v_{k_1}) & G_{\sigma}^0(v+\hat{x},v_0)
	\end{array}
	\right|
	\times
	\left|
	\begin{array}{ccc}
		G_{\bar{\sigma}}^0(v_1,v_1)-\frac{1}{2} & \cdots & G_{\bar{\sigma}}^0(v_1,v_{k_1})\\
        \vdots & \ddots & \vdots\\
	G_{\bar{\sigma}}^0(v_{k_1},v_1) & \cdots & G_{\bar{\sigma}}^0(v_{k_1},v_{k_1})-\frac{1}{2}\\
	\end{array}
	\right|~,
	\label{eqn:G1}
\end{equation}
while $G_2$ from $v$ to $v_0+\hat{x}$ or $v_0-\hat{x}$ with $V_2$ can be extracted from
\begin{equation}
	\left|
	\begin{array}{ccc|c}
		G_{\sigma}^0(v_{k_1+1},v_{k_1+1})-\frac{1}{2} & \cdots & G_{\sigma}^0(v_{k_1+1},v_k) & G_{\sigma}^0(v_{k_1+1},v)\\
		\vdots & \ddots & \vdots & \vdots \\
	G_{\sigma}^0(v_k,v_{k_1+1}) & \cdots & G_{\sigma}^0(v_k,v_k)-\frac{1}{2} & G_{\sigma}^0(v_k,v)\\
	\hline
	\makecell{G_{\sigma}^0(v_0+\hat{x},v_{k_1+1}) \\- G_{\sigma}^0(v_0-\hat{x},v_{k_1+1})} & \cdots & \makecell{G_{\sigma}^0(v_0+\hat{x},v_k) \\- G_{\sigma}^0(v_0-\hat{x},v_k)} & \makecell{G_{\sigma}^0(v_0+\hat{x},v) \\- G_{\sigma}^0(v_0-\hat{x},v)}\\
	\end{array}
	\right|
	\times
	\left|
	\begin{array}{ccc}
	    G_{\bar{\sigma}}^0(v_{k_1+1},v_{k_1+1})-\frac{1}{2} & \cdots & G_{\bar{\sigma}}^0(v_{k_1+1},v_k)\\
        \vdots & \ddots & \vdots\\
	G_{\bar{\sigma}}^0(v_k,v_{k_1+1}) & \cdots & G_{\bar{\sigma}}^0(v_k,v_k)-\frac{1}{2}\\
	\end{array}
	\right|~.
	\label{eqn:G2}
\end{equation}
The final bubble contribution will be the product of those two, $G_1\times G_2$~.
In our implementation, the Monte Carlo weight of a given $V$ configuration consists of the sum over all possible $(V_1,V_2)$ combinations. 
Since the CDet algorithm intrinsically computes the determinants of all subset matrices, we can construct $2^k$ different vertex distributions from a single calculation of the connected parts in Eq.~(\ref{eqn:sameSpin})~.
Furthermore, in this way, when there exist cancellations between different sign contributions from different combinations of $(V_1,V_2)$, we can enforce the cancellation of error among them.
It turns out that such cancellation indeed appears for our typical parameter sets.

\section{Series convergence}
\begin{figure}[h]
    \centering
    \includegraphics[width=1.0\textwidth]{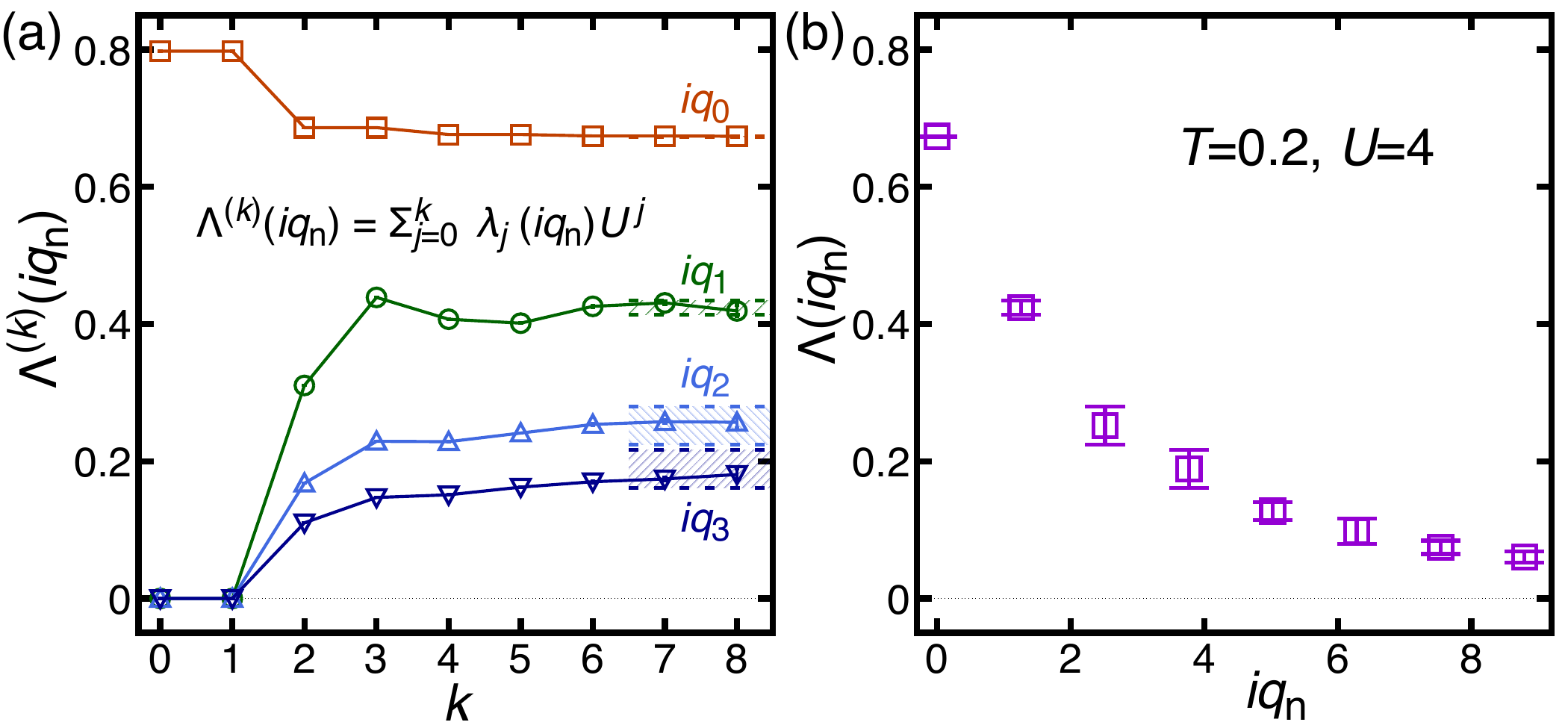}
    \caption{
	    (a) The partial sum of the diagrammatic series of the CCF for $T=0.2$ and $U=4$.
	    The results for the first four Matsubara frequencies are shown.
	    The shaded area for each Matsubara frequency presents the extrapolated error bounds.
	    (b) The extrapolated correlation function obtained from the series in the panel (a).
    }
    \label{fig:lmdan_series_set}
\end{figure}
As a result of DiagMC sampling, we obtain a set of diagrammatic series $\{\lambda_j(iq_n)\}$ for the current-current correlation function (CCF) of different Matsubara frequencies $\Lambda(iq_n)$.
Figure~\ref{fig:lmdan_series_set}(a) shows the partial sum of the series, $\Lambda^{(k)}(iq_n) = \sum^{k}_{j=0}\lambda_j(iq_n)U^j$, for the first four Matsubara frequencies up to the diagram order 8.
We typically use Pad\'e approximation for the extrapolation to infinite orders.
Even in challenging parameter regimes, e.g. $T=0.2$ and $U=4$, the diagrammatic series converge well, and the final results $\Lambda(iq_n)$ can be obtained within the controlled error bars, as shown in Fig.~\ref{fig:lmdan_series_set}(b)~.
The error bars shown include the fluctuation of the extrapolated value due to different error realizations in the series and different Pad\'e orders.

\section{Multiscale stretch test}
\begin{figure}[]
	\centering
	\includegraphics[width=0.9\textwidth]{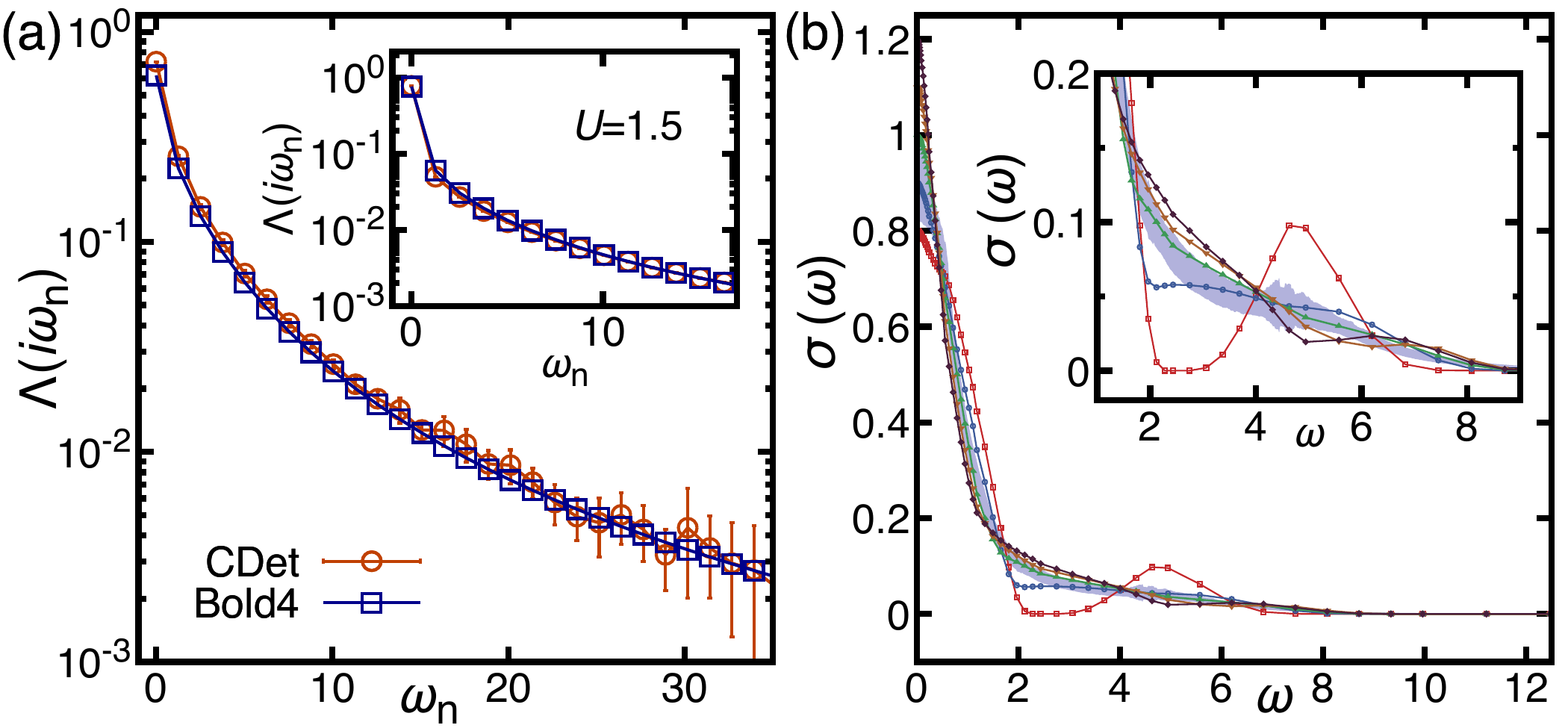}
	\caption{
		Multiscale stretch test for $T=0.2$ and $U=3.2$~. 
		(a) Matsubara CCF for $T=0.2$ and $U=3.2$ ($U=1.5$ for inset) obtained by the CDet (red circles) and Bold4 (blue squares) methods.
		(b) Optical conductivity obtained by the MCC with different target $\sigma(\omega=0)$ values: from top to bottom 1.2, 1.1, 1.0, 0.9, and 0.8 for $T=0.2$ and $U=3.2$~.
	}
	\label{fig:multiscaleNAC}
\end{figure}
As discussed in the main text, we perform the stretch test of the DC conductivity using the CCF whose high-frequency tail is substituted by the nonstochastic Bold4 results~\cite{Simkovic2017}.
Although missing high-order skeleton diagrams in the Bold4 scheme can make sizable effects in the low-frequency, it can provide the accurate, most importantly non-stochastic results for the high-frequency tail of the CCF as shown in Fig.~\ref{fig:multiscaleNAC}(a) inset.
For $T=0.2$ and $U=3.2$ [Fig.~\ref{fig:multiscaleNAC}(a)], for example, we substitute the CCF of the CDet with the one of the Bold4 for $\omega_n>15$~, where the error bar of the CDet becomes larger than the difference between them.
In order to take care of the potential systematic deviation from exact results, we introduce the $\mathcal{O}(1/\omega_n^3)$ error for the Bold4 tail.
It turns out that such tail substitution improves the quality of the stretch test.

Now we introduce the constraint which fixes the target $\sigma_{\text{DC}}$ value and see how the overall functional form of the optical conductivity changes using the method of consistent-constraints (MCC)~\cite{Prokofev2013}.
Under the MCC, a set of the spectral regularization parameters which enforce the smoothness and non-negativity and suppress deviations from the target DC conductivity, are applied and eventually reduced to small enough value to admit the solution of Eq.~(4) in the main text.
In this way, one can achieve the smooth spectra (free from notorious saw-tooth feature) without compromising the accurate fit of $\Lambda(iq_n)$ within the errorbars, i.e. $\chi^2\lesssim 1$ where
\begin{equation}
	\chi^2 = \sum^{L}_{j=0}\left[\frac{\Lambda(i\omega_j)-\tilde{\Lambda}(i\omega_j)}{\sigma_j}\right]^2~\quad\text{and}\quad \tilde{\Lambda}(i\omega_j) = \sum^{}_{k}K_{jk}\sigma(\omega_k)~.
	\label{eqn:chi}
\end{equation}

Note that the optical spectrum obtained by the MCC has strong mutual correlation between different frequency values; the local changes at $\omega=0$ can have a significant impact on entire functional form in finite frequencies.
And the range of $\sigma(\omega=0)$ which does not modify the overall functional form defines our error bar of the stretch test.
Figure~\ref{fig:multiscaleNAC}(b) presents the results for $T=0.2$ and $U=3.2$~.
One can see that $\sigma(0)=0.8$ line (red) generates the additional peak near $\omega=5$, which is absent in the base function (blue shade), so is clearly outside of the error bar. 
Even for $\sigma(0)=$0.9, 1.1, and 1.2~, the wiggly structure develops near $2\leq \omega\leq 8$ as shown in the inset of Fig.~\ref{fig:multiscaleNAC}(b), while the spectrum starting from the $\sigma(0)=$1.0 point nicely passes the stochastic optimization method (SOM) estimates for the full frequency range.
So our estimate of the DC conductivity $\sigma(0)\sim 1\pm 0.1$, which is consistent with the SOM result.

\section{Effect of van Hove Singularities on Charge Compressibility}
\begin{figure}[]
	\centering
	\includegraphics[width=0.5\textwidth]{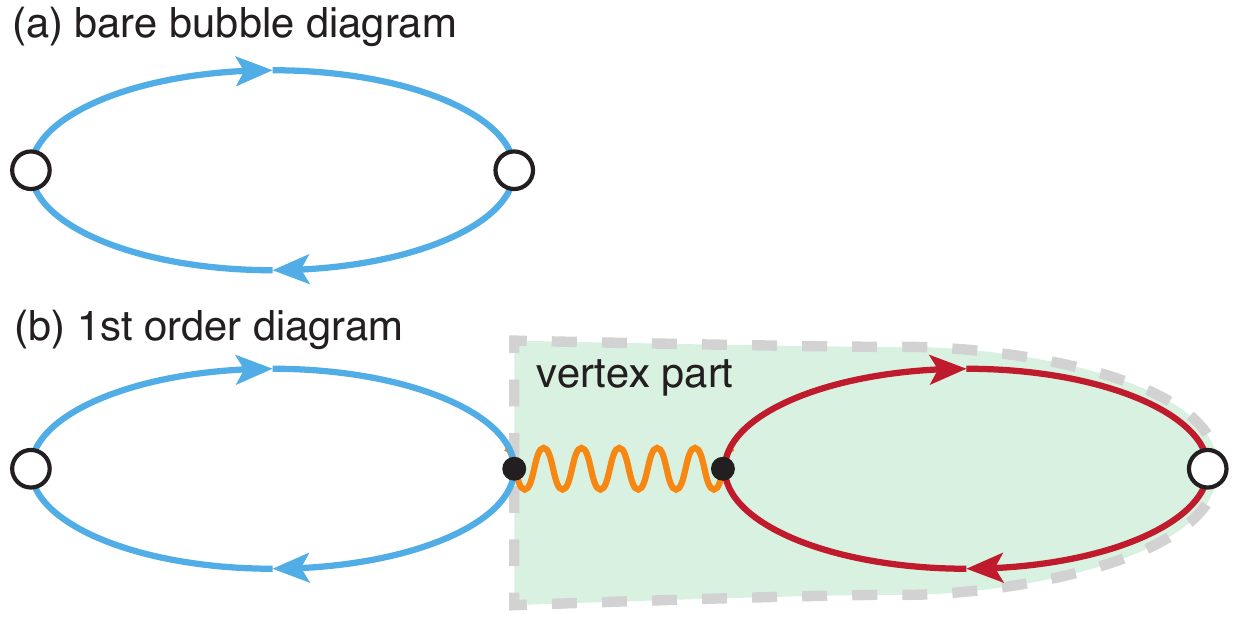}
	\caption{
		Diagrams for the compressibility: (a) the bare bubble and (b) the first-order diagrams. Blue and red lines denote the spin up and down components, respectively, and green shade represents the vertex part.
	}
	\label{fig:kappa_diagram}
\end{figure}

\begin{figure}[h]
    \centering
    \includegraphics[width=1.0\textwidth]{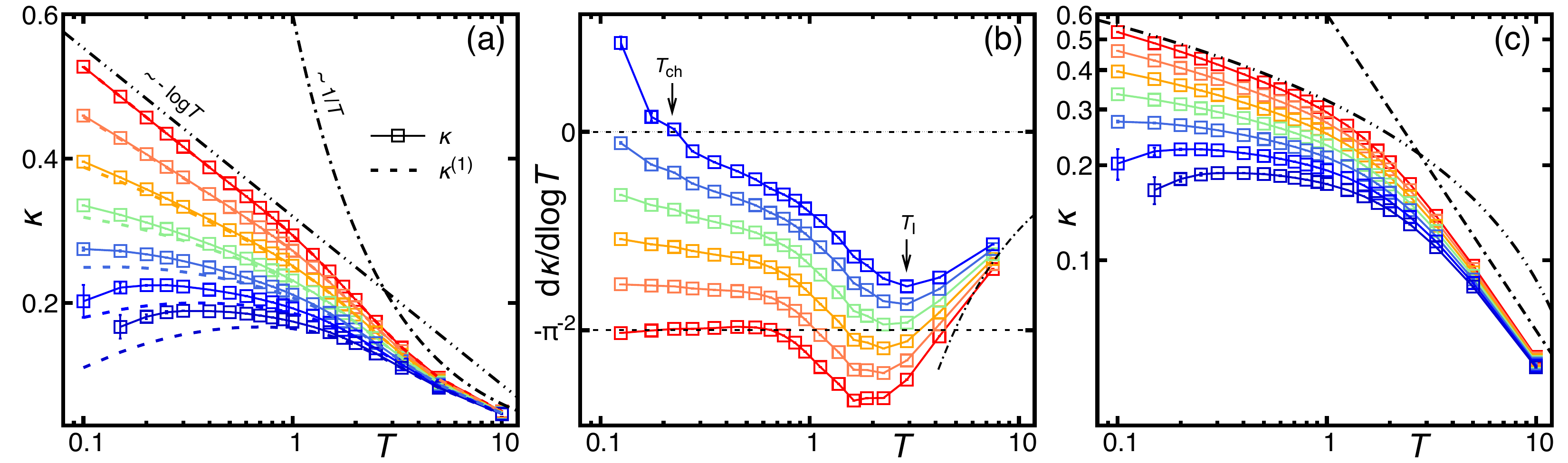}
    \caption{
	Compressibility in (a) the semi-log and (c) the log-log scales. From red to blue, the corresponding interaction strength $U=0.0, 0.5, 1.0, 1.5, 2.0, 2.5$, and $3.0$~.
	The middle panel (b) shows the (log) temperature derivative of the compressibility $d\kappa/\log T$, showing the first-order vertex correction in low temperatures. In low temperatures, the linear scaling appears while in high temperatures, the $1/T$ behavior of the compressibility is translated into the exponential decay (dot-dashed line) along the semi-log axis.
}
    \label{fig:vHS_kappa_set}
\end{figure}

In the noninteracting limit, the charge compressibility can be expressed by a simple bubble diagram [Fig.~\ref{fig:kappa_diagram}(a)], whose analytic expression reads
\begin{equation}
	\kappa = -\int_{}^{}d\epsilon~\frac{\partial n_F(\epsilon)}{\partial \epsilon} D(\epsilon)~,
	\label{<+label+>}
\end{equation}
where $n_F(\epsilon)$ is the Fermi function and $D(\epsilon)$ is the density of states.
The high-temperature behavior of the Fermi function, $\partial n_F(\epsilon)/\partial\epsilon\sim -\beta/4$, leads to the $1/T$ scaling of the compressibility while the $|\log\epsilon|$ dependence of $D(\epsilon)$ due to the van Hove singularity manifests itself when the peak of $-\partial n_F(\epsilon)/\partial\epsilon$ becomes sharp enough to resolve it at low temperatures.
The resulting logarithmic temperature dependence of the compressibility in low temperatures, $\kappa\sim |\log T|$, is one of the hallmarks of the van Hove singularity at the anti-nodal point of the Brillouin zone.
Figure~\ref{fig:vHS_kappa_set}(a) (red line) shows the high-to-low temperature crossover in the noninteracting limit: $|\log T|$ in low temperatures and $1/T$ in high temperatures.

When the Hubbard interaction is introduced, the first-order (in $U$) correction comes from the RPA-like (bubble) vertex [See Fig.~\ref{fig:kappa_diagram}(b)], which contributes the $-U(\log T)^2$ dependence of the compressibility in low temperatures.
As the temperature decreases, the $-U(\log T)^2$ contribution gets bigger compared to the bare counterpart ($|\log T|$) and eventually dominates the low temperature behavior, giving rise to the insulating compressibility $\partial\kappa/\partial T>0$~.
Considering only those two contributions, the metal-to-insulator crossover temperature scales exponentially as $T_{\text{ch}}\sim e^{-const/U}$ in the weak-coupling limit.
The dashed lines in Fig.~\ref{fig:vHS_kappa_set}(a) [$\kappa^{(1)}$ for various $U$ values] represent the first-order calculations, and they are qualitatively, even quantitatively in the low $U$ cases, consistent with the numerically exact high-order results.
The sign change of the compressibility slope corresponds to the metal-to-insulator crossover temperature denoted by $T_{\text{ch}}$ in the main text. 

The separation between the temperature scales (the van Hove singularity dominated low temperature and the ultra-high temperature) can be demonstrated more vividly in the $d\kappa/d\log T$ vs $\log T$ plot.
In Fig.~\ref{fig:vHS_kappa_set}(b), the derivative of the noninteracting compressibility is almost constant ($-\pi^{-2}$) at low temperatures.
The linear correction coming from the first-order bubble is clearly shown with the negative slope at low temperatures, while in the high-temperature regime, on the other hand, the derivative decays exponentially from the negative side.
Those two limiting behaviors inevitably indicate the existence of the derivative minimum(s) in between, which turns out to be a single minimum in all parameter sets we investigated, and we can also define this temperature ($T_{\text{I}}$) as a boundary between the thermal state and Diffusive Metal I.
Interestingly, this derivative minimum of the compressibility turns out to coincide with the temperature where the diffusion constant starts to increase.
We believe that such simultaneous changes in the qualitative behavior of $\kappa$ and the diffusion constant signal the complete transformation of the transport mechanism.

We also found yet another qualitative change in the compressibility between $T_{\text{ch}}$ and $T_{\text{I}}$.
As shown in Fig.~\ref{fig:vHS_kappa_set}(b), right below $T_\text{I}$, there is a clear temperature window, in which $d\kappa/d\log T$ shows convex curvature and the functional form clearly differs from that one of the first-order bubble above $T_{\mathrm{ch}}$~.
We define the temperature of the inflection point ($d^3\kappa/d\log T^3=0$) between $T_I$ and $T_{\mathrm{ch}}$ as $T_{\text{II}}$, separating the Diffusive Metal I ($T_\text{I}>T>T_\text{II}$) and II ($T_\text{II}>T>T_\text{ch}$)~.
Below $T_{\mathrm{II}}$, the system enters the low-temperature regime where the effect of the van Hove singularity becomes visible.
Technically, we use the 5th-order polynomial fit for $d\kappa/d\log T$ to find the inflection point.

\section{Vertex correction: second-order analytic expression and spin-dependent decomposition}
\begin{figure}[]
	\centering
	\includegraphics[width=0.5\textwidth]{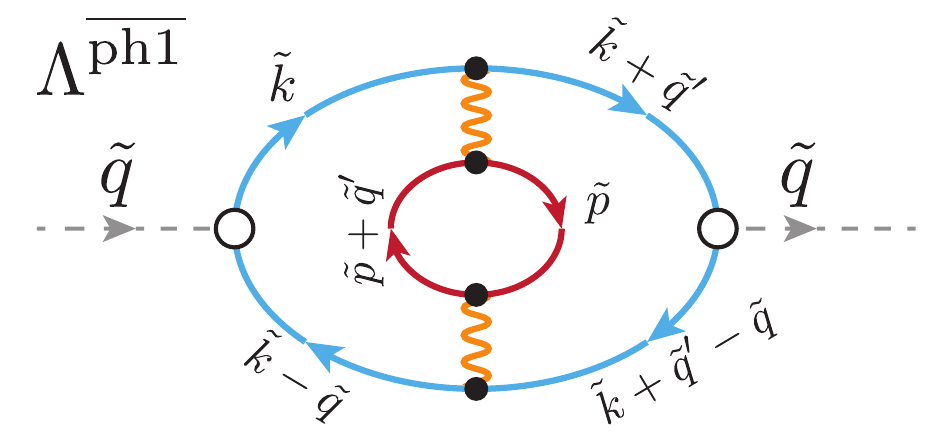}
	\caption{
		Only non-vanishing current vertex diagram at the second order. Blue (red) lines represent the noninteracting Green's function of spin $\uparrow$ ($\downarrow$), and white (black) dots present the external current (internal density) operators.
	}
	\label{fig:current_vertex}
\end{figure}

\begin{figure}[]
	\centering
	\includegraphics[width=0.9\textwidth]{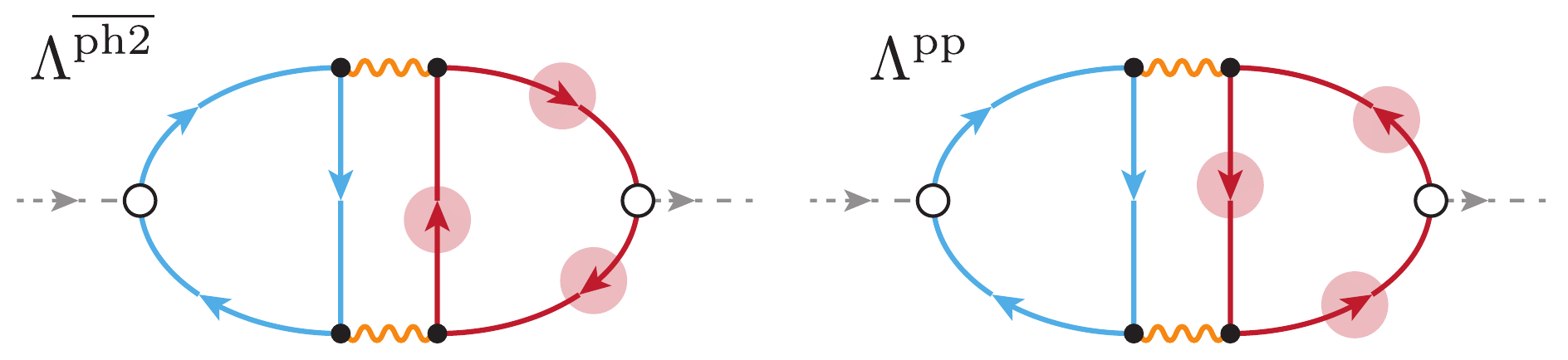}
	\caption{Opposite spin vertex correction at the second order. Two diagrams cancel each other.}
	\label{fig:vanishing2ndVertex}
\end{figure}

For the CCF, the first non-vanishing vertex correction appears at the second order.
There are three diagrams shown in the Fig.~\ref{fig:current_vertex} and \ref{fig:vanishing2ndVertex}, each of which is denoted by $\Lambda^{\overline{\text{ph1}}}$, $\Lambda^{\overline{\text{ph2}}}$, and $\Lambda^{\text{pp}}$; the overall vertex correction
\begin{equation}
	\Lambda^{(2)}_{vc}(\mathbf{q},iq_n) = \Lambda^{\overline{\mathrm{ph}1}}(\mathbf{q},iq_n)  + \Lambda^{\overline{\mathrm{ph}2}}(\mathbf{q},iq_n) + \Lambda^{\mathrm{pp}}(\mathbf{q},iq_n)~.
\end{equation}
The vertex symbols represent the 2-particle reducible channel for the 4-point vertex; the central block of $\overline{\text{ph}}$ ($\text{pp}$) vertex is reducible by cutting two horizontal lines of opposite (parallel) directions~\cite{Kauch2020}.

One can carry out the Matsubara sum over internal frequencies of those vertex diagrams. For example,
\begin{align}
	\Lambda^{\overline{\mathrm{ph}1}}(\mathbf{q}=0,iq_n) =& \frac{U^2}{(\beta N_x)^3}\sum_{\sigma}\sum_{\mathbf{k,p,q'}}v_\mathbf{k}v_{\mathbf{k+q'}}\sum_{ik_n,ip_n,iq'_n}\Big[G^0_{\sigma}(\mathbf{k},ik_n)G^0_{\sigma}(\mathbf{k+q'},ik_n+iq'_n)\nonumber\\
	&\times G^0_{\sigma}(\mathbf{k+q'},ik_n+iq'_n-iq_n)G^0_{\sigma}(\mathbf{k},ik_n-iq_n)G^0_{\bar{\sigma}}(\mathbf{p+q'},ip_n+iq'_n)G^0_{\bar{\sigma}}(\mathbf{p},ip_n)\Big]~,\nonumber\\
	=&\frac{2U^2}{N_x^3}\sum_{\sigma}\sum_{\mathbf{k,p,q'}}v_\mathbf{k} v_{\mathbf{k+q'}}\nonumber\\
	&\times \frac{[n_F(\epsilon_\mathbf{k})-n_F(\epsilon_{\mathbf{k+q'}})][n_F(\epsilon_\mathbf{p})-n_F(\epsilon_{\mathbf{p+q'}})][n_B(\epsilon_{\mathbf{k+q'}}-\epsilon_\mathbf{k})-n_B(\epsilon_{\mathbf{p+q'}}-\epsilon_\mathbf{p})]}{(\epsilon_{\mathbf{k+q'}}-\epsilon_\mathbf{k}-\epsilon_{\mathbf{p+q'}}+\epsilon_\mathbf{p})[(\epsilon_{\mathbf{k+q'}}-\epsilon_\mathbf{k}-\epsilon_{\mathbf{p+q'}}+\epsilon_\mathbf{p})^2 + q_n^2]}~,
\end{align}
where $v_{\mathbf{k}}=2t\sin k_x$ and $n_B$ is the Bose function~.

In the half-filled case with the particle-hole symmetry, there exists a symmetry relation of the noninteracting Green's function
\begin{equation}
	G^0(\mathbf{x},\tau) = -G^0(-\mathbf{x},-\tau)~,
	\label{eqn:G0sym}
\end{equation}
which leads to massive cancellations between existing diagrams.
Diagrammatically, Eq.~(\ref{eqn:G0sym}) implies that two $G^0$ lines running in opposite directions carry opposite signs of the same magnitude, so that every closed loop with an odd number of Green’s function lines is canceled by its opposite-direction counterpart.
This can be viewed as Furry's theorem for the Hubbard model, which is introduced in the context of quantum electrodynamics~\cite{Furry1937}.
Note that such diagrammatic relation holds for a specific choice of Hamiltonian decomposition. Even for the half-filled case, if the noninteracting part is shifted by the (generalized) alpha shift~\cite{Wu2017}, Eq.~(\ref{eqn:G0sym}) does not hold.

As an example of the diagrammatic cancellation in the second order, the two vertex diagrams, $\Lambda^{\overline{\text{ph2}}}$ and $\Lambda^{\text{pp}}$, cancel each other exactly leaving the $\Lambda^{\overline{\text{ph1}}}$ as the only non-vanishing one.
Note that all CCFs with opposite-spin vertex contributions vanish at the second order, and interestingly, this statement still holds for \textit{general even orders}.

\begin{figure}[]
	\centering
	\includegraphics[width=0.7\textwidth]{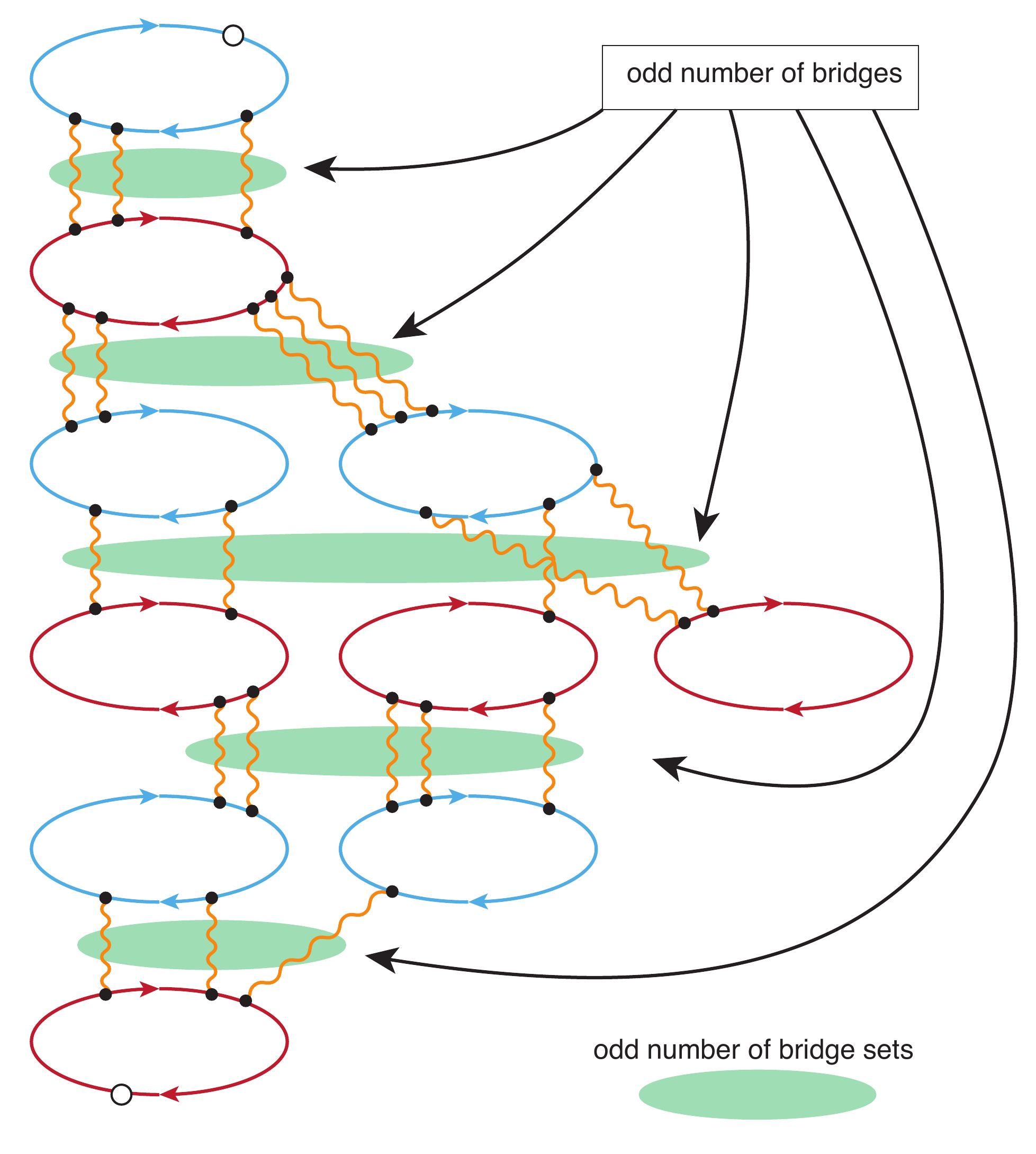}
	\caption{Valid bold-line loop diagrams with opposite spin external vertices. Note that the self-energy correction doesn't change evenness or oddness of the total diagram order.}
	\label{fig:high_order_proof}
\end{figure}
Suppose a diagram with two external vertices located in different loops, which should be the case for the opposite spin CCF.
Then one can reorganize the loops following the layered structure with an alternating spin sequence as in Fig.~\ref{fig:high_order_proof}.
Note that the self-energy correction of the Green's function doesn't change the evenness or the oddness of the diagram order since there are only nonvanishing self-energy at even orders at the half filling.
Since the top loop with a single external vertex should have even number of vertices in total not to be canceled, the number of internal vertices bridging the neighboring layer loop(s) should be an odd.
Again since the intermediate loops should have even number of vertices in total, if the number of incoming bridges to the layer is odd, the number of outgoing bridges is also odd. 
Finally, the number of boundaries between layers should be odd in order to achieve the opposite spin CCF as in Fig.~\ref{fig:high_order_proof} leading to the total odd number of bridges.
So, the opposite-spin CCF appears only at odd orders.

A Similar argument holds for the same-spin vertex, whose external vertices are located in different loops.
But this time, the total number of layer boundaries should be even in order to end up with same-spin loop.
So, the total diagram order will be even.

\begin{figure}[]
	\centering
	\includegraphics[width=0.5\textwidth]{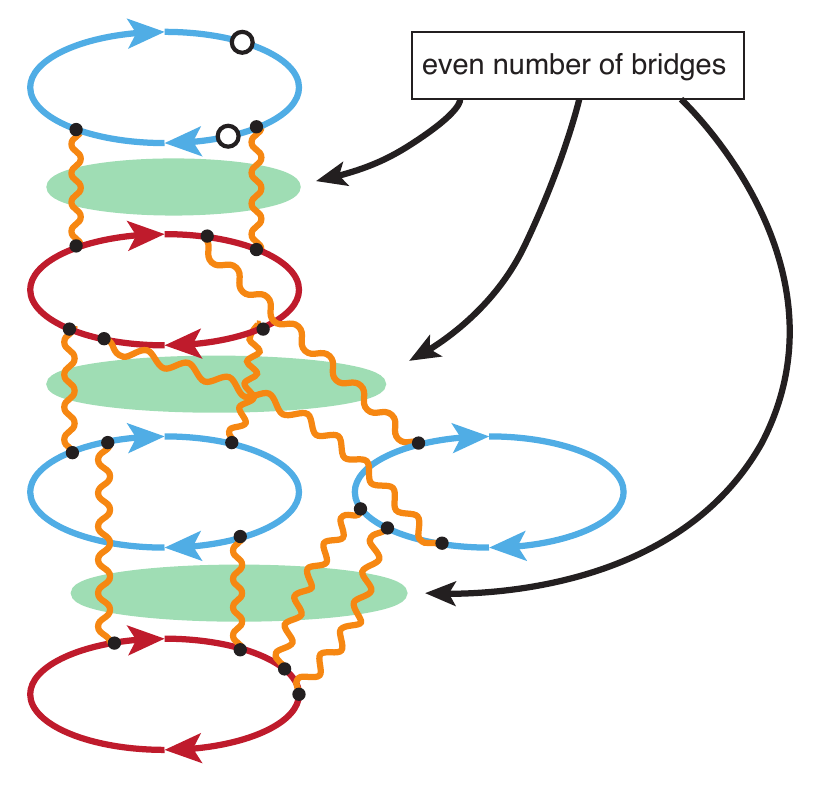}
	\caption{Example of the bold-line loop diagrams with same-spin external vertices in a single loop.}
	\label{fig:high_order_proof2}
\end{figure}
One can also consider the case where the two external vertices are located in the same loop; this includes generalized bubble diagrams.
When the two external vertices are in the same loop, the number of bridges connected to the neighboring layers should be even to make the total number of $G^0$ segments even [See Fig.~\ref{fig:high_order_proof2}]. 
Since the descendant layers have even number of incoming bridges, the number of outgoing bridges should also be even. 
So, independent of the number of layers, the total number of bridges should be even.

In conclusion, the vertex correction of the opposite-spin CCF will appear only in the odd orders while the bubble and vertex diagrams with the same-spin external vertices have even order contributions only.
Thus, by separately sampling the bubble diagrams at even orders, we can resolve the bubble and the spin-dependent vertex contributions.

\end{document}